\documentstyle[prd,aps,preprint]{revtex}
\tighten

\newcommand\Mpl{M_{\rm Pl}}

\def\be {\begin{equation}}
\def\ee {\end{equation}}
\begin{document}

\title{A New Mechanism for Baryogenesis in \\
Low Energy Supersymmetry Breaking Models}

\author{R. Brandenberger$^{1}$\footnote[1]{rhb@het.brown.edu.}
and
 A. Riotto$^{2}$\footnote[2]{riotto@nxth04.cern.ch, on leave of absence from Theoretical Physics Dept., Oxford Univ., Oxford, U.K.}}

\smallskip

\address{~\\$^1$Physics Department, Brown University, Providence, RI. 02912,
USA.}

\address{~\\$^2$Theory Group, CERN, CH-1211 Gen\'eve 23, Switzerland.}

\maketitle

\vskip 1.5cm
\begin{abstract}

\noindent A generic prediction of models where supersymmetry is broken at scales  within a few orders of magnitude of the weak scale and is fed down to the observable sector by gauge interactions  is the existence of superconducting cosmic strings which carry baryon number. In this paper we propose a novel mechanism for the generation of the baryon asymmetry which takes place at temperatures much lower than the weak scale. Superconducting strings act like ``bags'' containing the  baryon charge and protect 
it from sphaleron wash-out throughout the evolution of the Universe,  until
baryon number violating processes  become harmless. 
This mechanism is efficient  even if the electroweak phase transition in the MSSM  is of the second order
and  therefore does not impose any upper bound on the  mass of the  Higgs boson.

\end{abstract}

\vfill

\setcounter{page}{0}
\thispagestyle{empty}

\vfill

\noindent BROWN-HET-1111 \hfill        February 1998.

\noindent CERN-TH/98-38

\noindent OUTP-98-03-P

\noindent hep-ph/yymmnn \hfill Typeset in REV\TeX

\vfill\eject

\baselineskip 24pt plus 2pt minus 2pt

\section{Introduction}

Baryogenesis is an important channel which can relate particle physics beyond the standard model to data. The challenge for particle physics is to determine how the observed baryon to entropy ratio of $n_B / s \sim 10^{-10}$ can be generated from symmetric initial conditions ($n_B = 0$) in the very early Universe (here, $n_B$ and $s$ denote the net baryon number and entropy density, respectively).

For baryogenesis to be possible, it is necessary for the particle physics model to admit baryon number violating processes, to have $C$ and $CP$ violation, and for the relevant processes to occur out of thermal equilibrium. If $B-L$ is a symmetry of the theory, then \cite{KRZ} sphaleron processes which are unsuppressed above the electroweak symmetry breaking scale $\eta_{EW}$ will erase any baryon asymmetry generated at temperatures higher than $T_{EW}$, the temperature corresponding to $\eta_{EW}$, and the presently observed $n_B / s$ ratio must be generated below $T_{EW}$ (for a recent review of baryogenesis, see Ref. \cite{Dolgov}). 

A lot of recent work (see e.g. the review articles  \cite{Turokrev,CKNrev,Shaprev}) has focused on ways of generating the baryon asymmetry by electroweak sphaleron processes at or below $T_{EW}$. It is almost universally agreed (see, however, Ref. \cite{Farrar} for a dissenting view) that this is not possible in the standard model, firstly because $CP$ violation couples to sphaleron processes with insufficient strength, secondly because the electroweak phase transition is not strongly first order and hence does not proceed by the nucleation of critical bubbles. Such critical bubbles are the sites of non-equilibrium which are usually used for electroweak baryogenesis. The constraints of demanding that the phase transition be strongly first order, and that baryon nonconserving processes be out of equilibrium immediately below $T_{EW}$ are hard to reconcile even in extensions of the standard model. For instance, a successful electroweak baryogenesis within the Minimal 
Supersymmetric Standard Model (MSSM)   requires the Higgs boson and the right-handed stop to be  light and CP-violating phases  of the order of 0.1 \cite{carena}. 

There is, however, another mechanism which can be effective in mediating baryogenesis. Topological defects are out-of-equilibrium field configurations which can trigger baryogenesis in several ways: they can act as field condensates which release baryons during their decay  \cite{BDH}, or they can be used as the catalysis sites of sphaleron processes in a way similar to how bubble walls function \cite{BD,BDT,BDPT}.

In this paper, we work out an idea first suggested in \cite{Riotto97} that defects may play a crucial role in baryogenesis in a class of supersymmetric theories where supersymmetry breaking occurs at a low energy scale. It has been shown in previous work \cite{Riotto97} that these models admit superconducting cosmic strings with baryon number carrying condensates. 

As we will discuss, these strings are able to  trap the  baryon number from the time of string formation ($T = T_c$) until the temperature of the Universe has dropped down   to values much lower than  $T_{EW}$,  when sphaleron processes have fallen out of equilibrium. The generation of the baryon asymmetry takes place when the baryon number carried by the condensate in the core of the strings is released in the thermal bath at very low temperatures (this mechanism is different from defect-mediated electroweak baryogenesis \cite{BD,BDT,BDPT}, another mechanism which can be implemented in certain supersymmetric models \cite{TDB}). The  out-of-equilibrium condition is therefore naturally attained. 
 A  significant advantage of defect-mediated baryogenesis is  that the condition that sphaleron processes must be out of equilibrium immediately below $T_{EW}$ is no longer required {\cite{BDH}} and therefore  the stringent conditions on the MSSM Higgs spectrum may be relaxed.

Supersymmetric extensions are the best-motivated of the particle physics theories beyond the standard model. Supersymmetry solves the gauge hierarchy problem, couples gauge theory to gravity, and may easily generate the gauge coupling unification. However, the present state of the Universe is not supersymmetric, and thus supersymmetry must be broken in Nature at some scale higher than $\sim$ TeV.

Spontaneous supersymmetry breaking entails adding to the standard model and its supersymmetric copy some new fields acting as a supersymmetry breaking sector. The supersymmetry breaking sector cannot have any renormalizable couplings to the ``visible" sector, and is therefore often called ``hidden" or ``secluded". There are various ways in which supersymmetry breaking can be communicated from the ``hidden" to the ``visible" sector (see e.g. \cite{Dienes} for a recent review): supergravity mediation, gauge mediation, and mediation via $U(1)$ gauge factors. 
Since in the  most general  and natural  scenarios of low energy supersymmetry breaking   there are two different sources of contributions to ordinary superpartner masses,  gauge mediation from the messenger sector and  anomalous $U(1)$ $D$-terms,  a  novel feature is the  prediction of  superconducting  cosmic strings  \cite{Riotto97}. The typical energy scale of these strings is about $10^2$ TeV, and the bosonic charge carriers which render the strings superconducting are squark or slepton condensates. Hence, baryonic or leptonic charge may be stored in the core of the strings. After reviewing the particle physics model under consideration in Section 2, we study the evolution of the string network (Section 3). As we discover, the number density of strings at late times depends sensitively on whether the string loops become vortons or not. In Section 4, we calculate the baryon asymmetry per string loop decay, and use the result to estimate the overall $n_B / s$ ratio generate!
!
d by this mechanism.

\section{Low energy supersymmetry breaking models}

As we mentioned in the Introduction, supersymmetry  provides solutions  to many of the puzzles of the
standard model such as the stability of the weak scale under radiative
corrections as well as the origin of the weak scale itself. Since experimental observations require supersymmetry to
be broken, it is essential to have a knowledge of
the nature and the scale of supersymmetry breaking in order to have a
complete understanding of the physical implications of these theories.
At the moment, we lack such an understanding and therefore it is important 
to explore the various ways in which supersymmetry breaking can arise and
study their consequences. 

 The most  common and popular  approach is
to implement supersymmetry breaking in some  hidden
sector where some $F$-term gets a vacuum expectation value (VEV)  and then transmits it to the standard model sector  by
gravitational interactions. This is the so-called hidden $N=1$ supergravity scenario \cite{supersymmetry}. If one arranges the parameters in the hidden sector in such a way that the typical $\langle F\rangle $-term is of the order of $\langle F\rangle^{1/2} \sim\sqrt{(1\:{\rm TeV}) \Mpl}\sim 10^{11}$ GeV, where $\Mpl=2.4\times 10^{18}$ GeV,   the gravitino mass $m_{3/2}$ turns out to be of the order of  the TeV scale. In the $N=1$ supergravity scenario, however, the flavor-blindness is likely to be spoiled in the K\"{a}hler potential and one expects large contributions to the Flavour Changing Neutral Currents (FCNC's) at low energies. The suppression of the electric dipole moment of the neutron (EDMN) is also hard to explain in supergravity since  one expects the bilinear and trilinear terms to be of the order of $m_{3/2}$ and explicit CP-violating phases to be ${\cal O}(1)$.    

An alternative to hidden supergravity is provided by  the so-called gauge mediated supersymmetry breaking (GMSB) models \cite{gmsb}. The soft supersymmetry breaking mass terms of sfermions get  contributions at two-loop,   $\widetilde{m}^2\sim  
\left({{\alpha}\over{4\pi}}\right)^2\Lambda^2$, where $\Lambda\sim 10$ TeV  is the typical  scale of the messenger sector. The GMSB models have the extra advantage that
the FCNC effects are naturally suppressed. This is due to the fact that   
 at the scale $\Lambda$ the squark masses are
all degenerate because of  the flavor blindness of the standard model gauge group. Only a slight asymmetry is introduced by renormalization group extrapolation from the scale $\Lambda$ to low energies. Moreover, the trilinear soft breaking terms $A$ vanish at $\Lambda$ rendering the CP-violation problem
milder. 

Recently,  an alternative approach has been proposed in which gauginos, higgsinos and the third generation of squarks are sufficiently light to stabilize the electroweak scale, but the two first generations of squarks and sleptons  are sufficiently heavy to suppress the FCNC and the EDMN below the experimental bound \cite{pomarol,dudas,more}. This class of models is dubbed "more" minimally supersymmetric than the MSSM since they do not require some ad hoc supposition of degeneracy or alignment \cite{more}. In most of the attempts made so far along this line \cite{pomarol,dudas,riotto,nelson} 
the  crucial feature  is the
existence of an anomaly-free $U(1)$ gauge group, which appears anomalous below  some scale $M$. The effective theory includes a  Fayet-Iliopoulos (FI) $D$-term proportional to ${\rm Tr}\:Q M^2$ , where the  $Q$'s represent the $U(1)$ charges of the heavy fields which have been integrated out.  If the first two generations of squarks and sleptons, contrary to the third generation,  carry  $U(1)$ 
charges, the required mass hierarchy is obtained. 
We stress that supersymmetry models can be regarded as  realistic  only when they are   able to reproduce the quantitative features of the fermion spectrum and the CKM matrix. In a unified picture it is  desirable that the solution to all these puzzles  reside in the same sector of the theory. Recent investigations   have focused on the possibility that  a Frogatt-Nielsen flavor mechanism \cite{fn} is implemented
in the messenger sector \cite{riotto,nelson}. This means that the  messenger sector is also to be a Frogatt-Nielsen sector and that the 
messenger $U(1)$ symmetry by which the two first generation fermions get large masses is the same  Frogatt-Nielsen $U(1)$ for quarks and leptons.
Therefore, the most general (and maybe natural) scenario is an hybrid one  where  there are two different sources of contributions to ordinary superpartner masses,  gauge mediation from the messenger sector and  anomalous $U(1)$ $D$-terms \cite{riotto,nelson,inprep}.  

As shown in \cite{Riotto97}, this class of low energy supersymmetry breaking models naturally predicts (superconducting) cosmic strings. Indeed, the gauge group of the secluded sector,  where supersymmetry  is broken, may be  of  the form ${\cal G}\otimes U(1)_m$. The group $U(1)_m$ is usually some global $U(1)$  which is gauged and made anomaly-free.   
The fields  in the messenger sector  are charged under $U(1)_m$ and (some of them) under the standard model gauge group.   When supersymmetry is broken in the secluded sector,  some scalar fields in this sector  may acquire a VEV, but leave  the $U(1)_m$ gauge symmetry unbroken.  
The presence of an anomalous FI $D$-term $\xi$ may induce the spontaneous breakdown of the residual   $U(1)_m$ gauge symmetry  along some field direction in the messenger sector. In this case 
 local cosmic strings are formed \cite{Riotto97}. Indeed, integrating out the heavy fields belonging to the secluded sector amounts to generate a (say)  positive  one-loop  FI $D$-term $\xi$  and 
 negative two-loop soft supersymmetry  breaking squared masses  for (some of)  the scalar fields $\phi_i$ of the messenger sector. These squared masses are  generally proportional to the square of the corresponding  $U(1)_m$ charges $q_i$.  The
potential will read
\begin{equation}
\label{pot}
V=-\sum_iq_i^2\widetilde{m}^2\left|\phi_i\right|^2+\sum_i\left|\frac{\partial W}{\partial\phi_i}\right|^2+\frac{g^2}{2}\left(\sum_i q_i\left|\phi_i\right|^2+\xi\right)^2,
\end{equation}
where $g$ is the $U(1)_m$ gauge coupling constant.  From the minimization of (\ref{pot}), one can see that among the fields with $F$-flat directions, the one with the smallest negative charge (call it $\varphi$) will get a VEV
\begin{equation}
\langle|\varphi|^2\rangle=\frac{1}{q_\varphi}\left(q_\varphi\frac{\widetilde{m}^2}{g^2}-\xi\right).
\end{equation}
When this happens, the  residual global  $U(1)_m$ symmetry gets broken leading to the formation of   local strings whose mass per unit length is given by $\mu\sim\xi$. Since $\sqrt{\xi}$ is a few orders of magnitude larger than the weak scale, 
cosmic strings are not very heavy.  

We now  suppose that the quark and/or the lepton superfields are charged under the  $U(1)_m$ group.  As we argued, this is in the spirit of the "more" MSSM \cite{more} and somehow welcome when trying to quantitatively predict the fermion mass spectrum and the CKM matrix.  
Let us focus of one of the sfermion fields, $\widetilde{f}$ with $U(1)_m$-charge $q_f$ such that  sign $q_f=$ sign $q_\varphi$. The potential for the fields $\varphi$ and $\widetilde{f}$ is written as
\begin{equation}
V(\widetilde{f},\varphi)=-q_\varphi^2\widetilde{m}^2|\varphi|^2-q_f^2\widetilde{m}^2|\widetilde{f}|^2+
\frac{g^2}{2}\left(q_\varphi |\varphi|^2+q_f|\widetilde{f}|^2+\xi\right)^2+\lambda |\widetilde{f}|^4,
\end{equation}
where we have assumed, for simplicity, that $\widetilde{f}$ is $F$-flat. The parameter $\lambda$  is generated from the standard model gauge group $D$-terms and  vanishes if we take $\widetilde{f}$ to denote a family of fields parametrizing a $D$-flat direction. 

At the global minimum $\langle \widetilde{f}\rangle=0$ and the electric charge, the baryon and/or the lepton numbers are conserved. The soft breaking mass term for the sfermion reads
\begin{equation}
\Delta m_{\widetilde{f}}^2=q_f\left(q_\varphi-q_f\right)\widetilde{m}^2, 
\end{equation} 
which is positive in virtue of the hierarchy $q_\varphi<q_f<0$. Consistency with experimental bound requires $\Delta m^2_{\widetilde{f}}$ to be of the order of $\left(20\:{\rm  TeV}\right)^2$ or so,  which in turn requires $\xi\sim (4\pi/g^2)\widetilde{m}^2\sim (10^2\:{\rm TeV})^2$. Notice that $\Delta m_{\widetilde{f}}^2$ does not depend upon $\xi$. 

Let us analyse what happens in the core of the string. In this region of space, 
the vacuum expectation value of the field vanishes,   $\langle|\varphi|\rangle=0$,  and nonzero values of $\langle |\widetilde{f}|\rangle$ are energetically preferred in the string core 
\begin{equation}
\langle |\widetilde{f}|^2\rangle=\frac{\widetilde{m}^2 q^2_f-g^2\xi q_f}{g^2 q^2_f+2\lambda}.
\end{equation}
Since the vortex is cylindrically symmetric around the $z$-axis, the condensate will be of the form $\widetilde{f}=\widetilde{f}_0(r,\theta)\:{\rm e}^{i\eta_f(z,t)}$ where $r$ and $\theta$ are the polar coordinate in the $(x,y)$-plane. One can easily check that 
 the kinetic term for $\widetilde{f}$ also allows a nonzero value of $\widetilde{f}$ in the string and therefore one expects the existence of bosonic charge carriers inside the strings. The latter are, therefore, superconducting \cite{superc}.
 
What is noticeable is that, besides the electric charge, the baryon/and or the lepton numbers are also broken inside the string. In particular,   baryonic charge may be stored in the core of the string. Indeed, if the sfermion particles are identified with some squarks $\widetilde{q}$ (eventually parametrizing a standard model $D$-flat direction), they carry a $U(1)$ baryonic global charge which is derived from the conserved current
\begin{equation}
J^\mu_B=\frac{i}{2}\sum_q  q_B^q\left(\widetilde{q}^\dagger\partial^\mu\widetilde{q}-\widetilde{q}\partial^\mu\widetilde{q}^\dagger\right),
\end{equation}
where $q_B^q$ is the baryonic charge associated to any field $\widetilde{q}$. Under the assumption of cylindrical symmetry, the baryonic charge per unit length $Q_B$ along the $z$-axis will be given by $Q_B=\int d\theta dr\:r j_B(\theta,r)$ where $j_B$ is the current per unit length along the same axis.

\section{String Network Evolution}

In the class of models considered in Section 2, a network of strings forms during supersymmetry breaking at a temperature $T_c \sim 10^2$ TeV. String formation has been studied extensively \cite{Kibble,Zurek,HDB,BD2,Vachrev} both analytically and numerically. According to these studies, the initial string correlation length (mean separation between strings) is
\be \label{corrlength}
\xi(t_c) \sim \lambda^{-1} \eta^{-1} \, ,
\ee
$\eta\sim \sqrt{\xi}$ being the energy scale of the string, and $\lambda\sim g^2$ being a typical coupling constant of the $U(1)$ order parameter field. 

The evolution of a string network can be divided into two periods. In the initial period following the string-producing phase transition, the string dynamics is friction-dominated \cite{Kibble2,Everett,Hindmarsh}. During this period, the correlation length $\xi(t)$ increases super-luminally and catches up to the Hubble radius. The friction-dominated phase ends once $\xi(t) \sim t$. This happens at a time $t_f$ given by
\be
t_f \, \sim \, (G \mu)^{-1} t_c \, ,
\ee
$G$ being Newton's gravitational constant and $\mu = \eta^2$ the string mass per unit length. After $t_f$, the string network enters the scaling epoch during which $\xi(t) \sim t$. Since in our class of models $G \mu \sim 10^{-28}$, all of the physics relevant for baryogenesis takes place deep in the friction-dominated period.

It is important to know the rate at which $\xi(t)$ increases during the friction-dominated period. We take this rate to be given by
\be \label{corrincr}
\xi(t) \, = \, \xi(t_c) \left({t \over {t_c}}\right)^{3/2} \, .
\ee
This result can be justified as follows. For a string loop, tension dominates over friction if the loop radius $R$ is larger than a critical radius $r_f(t)$ which increases in time since the friction per unit string length decreases. A comparison between tension and friction gives \cite{Kibble2,Everett,Hindmarsh}
\be
r_f(t) \, = \, G \mu \Mpl^{1/2} t^{3/2}.
\ee
If $\xi(t)$ were smaller than $r_f(t)$, then the string network would evolve under the force of tension alone, and the resulting intercommutations of string would lead   to a rapid increase in $\xi(t)$ \cite{Vil85}. Conversely, if $\xi(t) \gg r_f(t)$, then strings would be approximately static in comoving coordinates, which would lead to a decrease of the ratio $\xi(t) / r_f(t)$. Hence,
\be 
\xi(t) \, \sim \, r_f(t) 
\ee
is a dynamical fixed point of the string network in the early period of evolution.

The rapid increase of $\xi(t)$ given by (\ref{corrincr}) is achieved by string loop production. Unless small-scale structure on the string network is important already in the friction-dominated epoch, the loops are produced with a radius proportional to $\xi(t)$ (see the reviews in \cite{HK,VS,RHBrev}). In this case, the number density of loops created per unit time is
\be \label{nodens}
{{dn} \over {dt}} \, = \, \nu \xi^{-4} {{d \xi} \over {dt}} \, ,
\ee
where $\nu$ is a constant of order unity.

In the class of models considered, the strings are superconducting and carry baryon charge. Let $Q_B$ denote the charge per unit length. By order of magnitude considerations, the value of $Q_B$ will be of the order $\eta$. Then, the charge on a correlation length of string at the time $t_c$ is
\be
Q_c \, = \, Q \xi(t_c) \, .
\ee
On longer pieces of string, the charge adds up as a random walk \cite{HE97,Enq,Vach}. Hence, loops created at time $t$ with radius $\xi(t)$ have a root mean square charge of
\be \label{chargeadd}
Q_B(t) \, = \, \left[{{\xi(t)} \over {\xi(t_c)}}\right]^{1/2} Q_c \, .
\ee

Whereas non-superconducting string loops decay only gravitationally with power $P_G = \gamma G \mu^2$, $\gamma$ being a constant whose value is about $100$ \cite{VS}, strings with a current on them decay predominantly by electromagnetic radiation. The power $P_\gamma$ of electromagnetic emission can be parametrized as \cite{OTW}
\be \label{emrad}
P_\gamma \, = \, \gamma_{em} j^2 \mu \, .
\ee
Here, $\gamma_{em}$ is a constant analogous to $\gamma$ (and for the numerical estamates which follow we will set them equal), and the relative current $j$ is defined by
\be
j = {{J} \over {J_{max}}} \, , \,\,\,\,\, J_{max} = e g_e^{-1} \eta \, ,
\ee
where $J$ denotes the current on the loop, and $g_e$ is the electromagnetic gauge coupling constant. Note that the current increases as the radius $R$ of the loop decreases from its initial value $R_0$:
\be \label{curincr}
j(R) \, = \, j_0 \left( {{R_0} \over R} \right) \, ,
\ee
$j_0$ being the initial relative current. Comparing the power of gravitational and electromagnetic radiation, it follows that if
\be \label{cond1}
j_0 \, > \, (G \mu)^{1/2} \, ,
\ee
then electromagnetic radiation is the dominant energy loss mechanism during the entire life-time of the loop. Note that
\be \label{raddec}
{\dot R} \, = \, (\beta \mu)^{-1} P \, ,
\ee
where $P$ is the total power of radiation and $\beta R$ is the length of the loop (in general, $\beta \neq 2 \pi$ if the loop is not a circle). Combining
(\ref{emrad}), (\ref{curincr}) and (\ref{raddec}), we can solve for the life-time of a string loop. We find that provided
\be \label{cond2}
j_0 \, > \, (G \mu)^{1/4} (R_0 \eta)^{1/2} \, ,
\ee
then the life-time of the loop is less than one Hubble expansion time.

The current per correlation length on the string is of the order $\eta$ (by dimensional arguments). On longer string segments, the current adds up as a random walk (analogous to (\ref{chargeadd})). Hence,
\be
j_0(R) \ \sim \, \left[ {R \over {\xi(t_c)}} \right]^{1/2} e g_e^{-1} \, ,
\ee
and hence the conditions (\ref{cond1}) and (\ref{cond2}) are clearly satisfied. Thus, on cosmological time-scales, superconducting strings instantaneously collapse. This fact will simplify the computation of the baryon asymmetry.

As mentioned in the Introduction, the cosmic string network provides a means for trapping baryon charge until late times. In the following section we will discuss how the charge trapped on a string loop gets converted into a net baryon number when the loop decays. For now we simply write
\be \label{asym}
\Delta n_B \, = \, Q_B \epsilon \, ,
\ee
for the net baryon number generated by the decay of a loop of charge $Q_B$, where $\epsilon$ is a constant related to the CP-violation parameter (see Section 4).

The resulting baryon asymmetry depends crucially on whether the loops decay immediately or collapse to form vortons \cite{vortons}, which themselves decay after $t_{EW}$. In the former case, the only loops which contribute to the net $n_B / s$ are those produced after $t_{EW}$:
\be \label{int1}
n_B(t) \, = \, \int_{t_{EW}}^t dt' \epsilon Q_B(t') {{dn} \over {dt}}(t') \left({{t'} \over t}\right)^{3/2} \, ,
\ee
where the final factor is due to the cosmological redshift. The integral (\ref{int1}) is dominated by the contribution from near $t_{EW}$ (or, more generally, the time when sphaleron processes fall out of equilibrium). Inserting (\ref{corrincr}), (\ref{nodens}) and (\ref{chargeadd}), the integral becomes
\be \label{barno1}
n_B(t) \, \simeq \, {2 \over 3} \epsilon \nu Q_c \xi(t_c)^{-3} \left({{t_c} \over t}\right)^{3/2} \left({{t_c} \over {t_{EW}}}\right)^{9/4} \, .
\ee
The last factor in (\ref{barno1}) is a geometrical suppression factor resulting since baryons produced by loops which decay before $t_{EW}$ equilibrate by sphaleron processes.

On the other hand, if superconducting loops form vortons which decay only after $t_{EW}$, then all loops created since $t_c$ contribute to the net $n_B$ which is determined now by an integral like (\ref{int1}) but with lower integration limit $t_c$. Since the integral is once again dominated by the lower limit we obtain
\be \label{barno2}
n_B(t) \, \simeq \, \epsilon \nu Q_c \xi(t_c)^{-3} \left({{t_c} \over t}\right)^{3/2} \, .
\ee 
The geometric suppression factor which appeared in (\ref{barno1}) is no longer present. For $T_c \sim 10^2$ TeV, baryogenesis is more efficient by a factor of $(T_c / T_{EW})^{4.5} \sim 10^9$ if the loops form vortons which are stable until at least $t_{EW}$. Obviously, the overall strength of our baryogenesis mechanism depends on the value of $\epsilon$, to the evaluation of which we now turn.

\section{The baryogenesis Mechanism}

As mentioned in the Introduction and discussed in Section 2, the strings in our class of theories carry baryonic charge in the form of a squark or slepton condensate. When the string loops decay, a net baryon charge is generated analogously to the Affleck-Dine mechanism for baryon number production \cite{ad}. We should comment here on two crucial features of the mechanism. First, when superconducting strings are formed, one may expect that the number of strings
with some baryonic charge $Q_B$ is equal to the number of strings with opposite
baryonic charge $-Q_B$. Therefore, when string loops decay and release the baryon number in the thermal bath, the net baryon asymmetry vanishes unless some source of {\it explicit} CP-violation is present  
in the interactions. Secondly, even though superconducting strings are formed at the temperature $T_c\gg T_{EW}$, the baryon charge stored in the core is not washed out by baryon number violating processes induced by sphaleron transitions. Indeed, sphalerons are not  active in the strings. This is because {\it any} condensate $\phi_i$ carrying
$SU(2)_L$ quantum numbers contributes to the sphaleron energy, $E_{{\rm sph}}\propto\sqrt{\sum_i|\phi_i|^2}$. Therefore, if the sfermion condensate in the core of the string is charged under $SU(2)_L$, sphaleron transitions are suppressed inside the strings (see also \cite{Perkins1}). This is equivalent to saying that the superconducting strings act like ``bags'' containing  baryon charge and protect 
it from sphaleron wash-out throughout the evolution of the Universe until
 baryon number violating processes  are rendered harmless. 

For sake of concreteness, let consider the case in which the sfermion condensate is formed by the third family left-handed stop $\widetilde{t}_L$ (one can easily generalize this case). The left-handed stop decay channels most important for our considerations are   $\widetilde{t}_L\rightarrow t_R+
\widetilde{H}_2^0$, $ H^0_2+\widetilde{t}_R$
where $H_2^0$, $\widetilde{H}_2^0$ and $\widetilde{t}_R$ are the Higgs,  the Higgsino and the right-handed stop,  respectively. The constant $\epsilon$ (introduced in (\ref{asym})) which determines the asymmetry in baryon and antibaryon production is determined by the difference in the decay widths between $\widetilde{t}_L$ and $\widetilde{t}_L^*$:
\be
\epsilon=\frac{\Gamma_{\widetilde{t}_L}-\Gamma_{\widetilde{t}_L^*}}{\Gamma_{\widetilde{t}_L}+\Gamma_{\widetilde{t}_L^*}}.
\ee
The value of $\epsilon$ is determined by the strength of the explicit CP-violation which couples to the $\widetilde{t}_L$ decay processes. For us, the dominant contribution is the CP-violating phase of the coefficient $A_t$ of the term
\be
A_t h_t \widetilde{t}_L \widetilde{t}_R^* H_0^2+{\rm h.c.}
\ee
in the interaction Lagrangian, where $h_t$ is top Yukawa coupling.

The contribution to $\epsilon$ comes from the interference terms between the tree level and one loop decay diagrams. Assuming that $m_t \simeq \widetilde{m}$ (the supersymmetry breaking mass) we obtain
\be
\epsilon \, \simeq \, {{{\rm ln} \:2} \over {8 \pi}}\:  h_t^2\: 
{\rm sin}\phi_{A_t},
\ee
where we have written $A_t=|A_t|\:{\rm e}^{i \phi_{A_t}}$. Notice that $\phi_{A_t}$ is not constrained from the experimental upper bound on the EDMN and therefore can be as large as unity. 

At this point, we can insert the value of $\epsilon$ obtained above into the general expressions for the baryon number obtained in Section 3. Replacing time $t$ by the corresponding temperature $T$ by means of the Friedmann equations, and making use of the expression for the entropy density $s$ in terms of the number of degrees of freedom $g_*$ in thermal equilibrium at $T_{EW}$
\be
s(T) \, \sim \, g_* T^3 \, ,
\ee
we obtain from (\ref{barno1}) and (\ref{corrlength}):
\be
{{n_B} \over s} \, \sim \, \epsilon \nu Q_c \lambda^3 (g_*)^{-1} \left({{T_{EW}} \over {T_c}}\right)^{4.5} 
\ee
for the case when the string loops collapse and disappear immediately, and from (\ref{barno2}) and (\ref{corrlength})
\be
{{n_B} \over s} \, \sim \, \epsilon \nu Q_c \lambda^3 (g_*)^{-1}
\ee
for the case when stable vortons form during the initial loop collapse and survive until after $t_{EW}$.

We conclude that in the case in which stable vortons form upon the collapse of a string loop, it is possible to obtain the observed value of $n_B / s$ without unnatural constraints on the coupling constants. To get the feeling with the numbers, for $\epsilon\sim 10^{-2}$, $\lambda\sim 10^{-1}$, $\nu \sim 1$ and $g_*\sim 10^2$, we need $Q_c\sim 10^{-3}$ to produce $n_B / s \sim 10^{-10}$. 
  
\section{Conclusions}

In this paper we have proposed a new mechanism for the generation of the baryon asymmetry in the early Universe. It is based on the general observation that, if  superconducting cosmic strings  form  at scales within a few orders of magnitude of the weak scale and carry some baryon charge, the latter is efficiently  preserved from the sphaleron erasure and may be released in the thermal bath  at low temperatures. A natural framework for this mechanism is represented by the class of models where supersymmetry is broken at low energy. In such  a case, the charge carriers inside the strings are provided by the scalar superpartner of the fermions which carry baryon (lepton) number. Since these scalar condensates are charged under $SU(2)_L$, baryon number violating processes are frozen in the core of the strings and the baryon charge number can not  be  wiped out at temperatures larger than $T_{EW}$. The mechanism is very efficient if stable vortons form upon the collapse of the string loops and survive until after the electroweak phase transition (note that the role of vortons in baryogenesis has recently also been studied in \cite{Perkins2}). 

We like to stress that our proposal has a number of advantages with respect to the idea of bubble-mediated electroweak baryogenesis since it does not ask for a sufficiently
strong electroweak first order phase transition and therefore does not impose any stringent bound on the mass spectrum of the MSSM. An important issue for future study is to investigate the probability of formation of vortons and their stability. As has been shown recently \cite{vortonbound}, cosmological constraints from the density parameter at the present time and from nucleosynthesis are consistent with vorton formation at the scale considered in this paper.

\centerline{\bf Acknowledgments}

This work is supported in part by the US Department of Energy under
contract DE-FG0291ER40688, Task A. R.B. wishes to thank Profs. W. Unruh and
A. Zhitnitsky for hospitality at the University of British Columbia, where some of the work was completed.

\def\NPB#1#2#3{{\it Nucl. Phys.} {\bf B#1}, #3 (19#2)}
\def\PLB#1#2#3{{\it Phys. Lett.} {\bf B#1}, #3 (19#2) }
\def\PLBold#1#2#3{{\it Phys. Lett.} {\bf#1B} (19#2) #3}
\def\PRD#1#2#3{{\it Phys. Rev. }{\bf D#1}, #3 (19#2) }
\def\PRL#1#2#3{{\it Phys. Rev. Lett.} {\bf#1} (19#2) #3}
\def\PRT#1#2#3{Phys. Rep. {\bf#1} (19#2) #3}
\def\ARAA#1#2#3{Ann. Rev. Astron. Astrophys. {\bf#1} (19#2) #3}
\def\ARNP#1#2#3{Ann. Rev. Nucl. Part. Sci. {\bf#1} (19#2) #3}
\def\MPL#1#2#3{Mod. Phys. Lett. {\bf #1} (19#2) #3}
\def\ZPC#1#2#3{Zeit. f\"ur Physik {\bf C#1} (19#2) #3}
\def\APJ#1#2#3{Ap. J. {\bf #1} (19#2) #3}
\def\AP#1#2#3{{Ann. Phys. } {\bf #1} (19#2) #3}
\def\RMP#1#2#3{{Rev. Mod. Phys. } {\bf #1} (19#2) #3}
\def\CMP#1#2#3{{Comm. Math. Phys. } {\bf #1} (19#2) #3}


\begin{thebibliography}{30}

\bibitem{KRZ} V. Kuzmin, V. Rubakov and M. Shaposhnikov, {\it Phys. Lett.} {\bf B155}, 36 (1985).
\bibitem{Dolgov} A. Dolgov, {\it Phys. Rep.} {\bf 222}, 309 (1992).
\bibitem{Turokrev} N. Turok, in `Perspectives on Higgs Physics', ed. G. Kane (World Scientific, Singapore, 1992).
\bibitem{CKNrev} A. Cohen, D. Kaplan and A. Nelson, {\it Ann. Rev. Nucl. Part. Sci.} {\bf 43}, 27 (1993).
\bibitem{Shaprev} V. Rubakov and M. Shaposhnikov, {\it Phys. Usp.} {\bf 39}, 461 (1996).
\bibitem{Farrar} G. Farrar and M. Shaposhnikov, {\it Phys. Rev. Lett.} {\bf 70}, 2833 (1993).
\bibitem{carena} M. Carena, M. Quiros, A. Riotto, I. Vilja and C.E.M. Wagner,  {\it Nucl. Phys.} {\bf B503}, 387 (1997). 
\bibitem{BDH} R. Brandenberger, A.-C. Davis and M. Hindmarsh, {\it Phys. Lett.} {\bf B263}, 239 (1991).
\bibitem{BD} R. Brandenberger and A.-C. Davis, {\it Phys. Lett.} {\bf B308}, 79 (1993).
\bibitem{BDT} R. Brandenberger, A.-C. Davis and M. Trodden, {\it Phys. Lett.} {\bf B332}, 305 (1994).
\bibitem{BDPT} R. Brandenberger, A.-C. Davis, T. Prokopec and M. Trodden, {\it Phys. Rev.} {\bf D53}, 4257 (1996).
\bibitem{Riotto97} A. Riotto, {\it Phys. Lett.} {\bf B413}, 22 (1997).
\bibitem{TDB} M. Trodden, A.-C. Davis and R. Brandenberger, {\it Phys. Lett.} {\bf B349}, 131 (1995).
\bibitem{Dienes} K. Dienes and C. Kolda, ``20 Open Questions in Supersymmetric Particle Physics", hep-ph/9712322.
\bibitem{supersymmetry} For reviews of supersymmetry and supergravity, see
        H. P. Nilles, {\it Phys. Rept.} {\bf 110}, 1 (1984); H.E. Haber and G.L. Kane, {\it Phys. Rept}. {\bf 117}, 75 (1985)
\bibitem{gmsb} The idea of gauge mediated supersymmetry breaking was originally discussed in M.~Dine, W.~Fischler, and M.~Srednicki, \NPB{189}{81}{575};
S.~Dimopoulos and S.~Raby, \NPB{192}{81}{353};
M.~Dine and W.~Fischler, \PLB{110}{82}{227};
M.~Dine and M.~Srednicki, \NPB{202}{82}{238};
M.~Dine and W.~Fischler, \NPB{204}{82}{346};
L.~Alvarez-Gaum\'e, M.~Claudson, and M.~Wise, \NPB{207}{82}{96};
C.R.~Nappi and B.A.~Ovrut, \PLB{113}{82}{175};
S.~Dimopoulos and S.~Raby, \NPB{219}{83}{479}.
 For a recent list of refs., see C. Kolda, hep-ph/9707450; G.F. Giudice and R. Rattazzi, hep-ph/9801271. 
\bibitem{pomarol} G. Dvali and A. Pomarol, {\it Phys. Rev. Lett.} {\bf 77}, 3728 (1996).
\bibitem{dudas} P. Binetruy and E. Dudas, {\it Phys. Lett.} {\bf B389},  503 (1996).
\bibitem{more} A.G. Cohen, D.B. Kaplan and A.E. Nelson, {\it Phys. Lett.} {\bf B388}, 588 (1996).
\bibitem{riotto} R. N. Mohapatra and A. Riotto, {\it Phys. Rev.} {\bf D55}, 1138 (1997); {\it Phys. Rev.} {\bf D55},  4262 (1997).
\bibitem{nelson} A.E. Nelson and D. Wright,  {\it Phys. Rev.} {\bf D56}, 1598 (1997). 
\bibitem{fn} C.D. Frogatt and H.B. Nielsen, {\it Nucl. Phys.} {\bf B147}, 277 (1979). 
\bibitem{inprep} D.E. Kaplan, F. Lepeintre, A. Masiero, A.E. Nelson and  A. Riotto, in preparation. 
\bibitem{superc} E. Witten, {\it Nucl. Phys.} {\bf B249}, 557 (1985). 
\bibitem{Kibble} T.W.B. Kibble, {\it J. Phys.} {\bf A9}, 1387 (1976).
\bibitem{Zurek} W. Zurek, {\it Acta Phys. Pol.} {\bf B24}, 1301 (1993).
\bibitem{HDB} M. Hindmarsh, A.-C. Davis and R. Brandenberger, {\it Phys. Rev.}
{\bf D49}, 1944 (1994).
\bibitem{BD2} R. Brandenberger and A.-C. Davis, {\it Phys. Lett.} {\bf B332}, 305 (1994).
\bibitem{Vachrev} T. Vachaspati, ``Formation of Topological Defects", hep-ph/9710292.
\bibitem{Kibble2} T.W.B. Kibble, {\it Acta Physica Polonica} {\bf B13}, 723
(1982).
\bibitem{Everett} A. Everett, {\it Phys. Rev.} {\bf D24}, 858 (1981).
\bibitem{Hindmarsh} M. Hindmarsh, PhD thesis, Imperial College (unpublished) (1986).
\bibitem{Vil85} A. Vilenkin, {\it Phys. Rept.} {\bf 121}, 263 (1985).
\bibitem{HK} M. Hindmarsh and T.W.B. Kibble, {\it Rept. Prog. Phys.} {\bf 58}, 477 (1995).
\bibitem{VS} A. Vilenkin and E.P.S. Shellard, {\it Cosmic strings and
other topological defects} (Cambridge Univ. Press, Cambridge, 1994).
\bibitem{RHBrev} R. Brandenberger, {\it Int. J. Mod. Phys.} {\bf A9}, 2117 (1994).
\bibitem{HE97} M. Hindmarsh and A. Everett, ``Magnetic Fields from Phase Transitions", astro-ph/9708004.
\bibitem{Enq} K. Enqvist and P. Olesen, {\it Phys. Lett.} {\bf B319}, 178 (1993).
\bibitem{Vach} T. Vachaspati, {\it Phys. Lett.} {\bf B265}, 258 (1991).
\bibitem{OTW} J. Ostriker, C. Thompson and E. Witten, {\it Phys. Lett.} {\bf B180}, 231 (1986).
\bibitem{vortons} R. Davis and E.P.S. Shellard, {\it Phys. Rev.} {\bf D38}, 4722 (1988); {\it Nucl. Phys.} {\bf B323}, 209 (1989); for a recent review see e.g. B. Carter, ``Recent Developments in Vorton Theory", astro-ph/9712116. 
\bibitem{ad} I. Affleck and  M. Dine, {\it  Nucl. Phys.}, {\bf B249}, 361 (1985).
\bibitem{Perkins1} W. Perkins, {\it Nucl. Phys.} {\bf B449}, 265 (1995).
\bibitem{Perkins2} A.-C. Davis and W. Perkins, {\it Phys. Lett.} {\bf B393}, 46 (1993). 
\bibitem{vortonbound} R. Brandenberger, B. Carter, A.-C. Davis and M. Trodden, {\it Phys. Rev.} {\bf D54}, 6059 (1996).

 

\end{thebibliography}
\end{document}